\documentclass[preprint,showpacs,superscriptaddress,preprintnumbers,amsmath,amssymb]{revtex4}

\usepackage[pdftex]{graphicx}
\usepackage{dcolumn}
\usepackage{bm}
\usepackage[thinspace]{SIunits}

\begin{document}


\title{Optimisation of NbN thin films on GaAs substrates for in-situ single photon detection in structured photonic devices}

\author{G. Reithmaier}
 \email{guenther.reithmaier@wsi.tum.de}
 \affiliation{Walter Schottky Institut, Technische Universit\"at M\"unchen, Am Coulombwall 4, 85748 Garching, Germany \\}
\author{J. Senf}
 \affiliation{Walter Schottky Institut, Technische Universit\"at M\"unchen, Am Coulombwall 4, 85748 Garching, Germany \\}
\author{S. Lichtmannecker}
 \affiliation{Walter Schottky Institut, Technische Universit\"at M\"unchen, Am Coulombwall 4, 85748 Garching, Germany \\}
\author{T. Reichert}
 \affiliation{Walter Schottky Institut, Technische Universit\"at M\"unchen, Am Coulombwall 4, 85748 Garching, Germany \\} 
\author{F. Flassig}
 \affiliation{Walter Schottky Institut, Technische Universit\"at M\"unchen, Am Coulombwall 4, 85748 Garching, Germany \\}
\author{A. Voss}
 \affiliation{Walter Schottky Institut, Technische Universit\"at M\"unchen, Am Coulombwall 4, 85748 Garching, Germany \\}
\author{R. Gross}
 \affiliation{Walther Mei\ss ner Institut, Technische Universit\"at M\"unchen,  Walther-Mei\ss ner-Stra\ss e 8, 85748 Garching, Germany \\}
\author{J.J. Finley}
 \email{j.j.finley@wsi.tum.de}
 \affiliation{Walter Schottky Institut, Technische Universit\"at M\"unchen, Am Coulombwall 4, 85748 Garching, Germany \\}

\date{\today}

\begin{abstract}
We prepare NbN thin films by DC magnetron sputtering on $[100]$ GaAs substrates, optimise their quality and demonstrate their use for efficient single photon detection in the near-infrared. The interrelation between the Nb:N content, growth temperature and crystal quality is established for $4-22$nm thick films. Optimised films exhibit a superconducting critical temperature of $12.6\pm0.2$K for a film thickness of $22\pm0.5$nm and $10.2\pm0.2$K for $4\pm0.5$nm thick films that are suitable for single photon detection. The optimum growth temperature is shown to be $\sim475^\circ{}$C reflecting a trade-off between enhanced surface diffusion, which improves the crystal quality, and arsenic evaporation from the GaAs substrate. Analysis of the elemental composition of the films provides strong evidence that the $\delta$-phase of NbN is formed in optimised samples, controlled primarily via the nitrogen partial pressure during growth. By patterning optimum $4$nm and $22$nm thick films into a $100$nm wide, $369\text{$\micro$m}$ long nanowire meander using electron beam lithography and reactive ion etching, we fabricated single photon detectors on GaAs substrates. Time-resolved studies of the photo-response, absolute detection efficiency and dark count rates of these detectors as a function of the bias current reveal maximum single photon detection efficiencies as high as $21\pm2$\% at $4.3\pm0.1$K with $\sim50$k dark counts per second for bias currents of $98\%I_\text{C}$ at a wavelength of $950$nm. As expected, similar detectors fabricated from $22$nm thick films exhibit much lower efficiencies ($0.004$\%) with very low dark count rates $\leq 3$cps. The maximum lateral extension of a photo-generated hotspot is estimated to be $30\pm 8$nm, clearly identifying the low detection efficiency and dark count rate of the thick film detectors as arising from hotspot cooling via the heat reservoir provided by the NbN film.

\end{abstract}

\pacs{74.78.-w 74.25.Gz 78.67.Uh 85.25.Oj 85.25.-j 42.50.-p}

\keywords{single photon detection, superconducting photodetectors, nanowires, thin film growth, quantum optics, GaAs, NbN, photonic devices}

\maketitle

Optical quantum information technologies \cite{OBrien09} would profit significiantly from the availability of single photon sources and detectors that can be compactly and directly integrated \emph{on-chip} together with passive optical components. In these respects, semiconductor based 2D photonic crystals (PCs) \cite{Krauss99} are highly attractive since quantum dots (QDs) serving as deterministic single photon sources can be embedded into them \cite{Kress05} and passive optical hardware such as waveguides \cite{Olivier02,McNab03,Notomi04}, beamsplitters \cite{Boscolo02} and phase-shifters can be integrated around them \cite{Fushman08}.  Recently, the efficient directional guiding of single photons from QDs into propagating PC waveguide modes has been theoretically \cite{MangaRao} and experimentally \cite{Schwagmann11,Laucht12} investigated, demonstrating the broad feasibility of this approach. Furthermore, several experiments have confirmed that the strong light-matter coupling regime is accessible in high-Q PC nanocavities containing individual QDs \cite{Laucht09,Ota11} illustrating that few photon quantum non-linearities could be exploited to provide the phase and amplitude control \cite{Fushman08} needed for photon based quantum information processing \cite{OBrien03}. Whilst much progress has been made, the on-chip \emph{detection} of single photons on photonic crystal nanostructures \cite{Heeres10} would complete the pallete of components needed for photon based quantum information technologies \cite{Knill01} in a highly integrated geometry.

Nanowire based superconducting single photon detectors (SSPDs) combine high detection efficiency, low dark count rates and picosecond timing resolution \cite{Goltsman01,Najafi12}, properties that strongly depend on the crystal quality of the NbN film and the fabrication process used to pattern the film \cite{Marsili11}. The growth of NbN thin films on MgO and sapphire substrates is well established \cite{Chockalingam08}, giving rise to critical temperatures as high as $T_\text{C}\sim16.4$K \cite{Treece95}. However, the growth of NbN on crystalline GaAs is challenging due to the $27\%$ lattice mismatch and the need to deposit films at low temperatures \cite{Marsili09} below the threshold for thermal decomposition. The figures of merit that determine SSPD operation are a high superconducting critical temperature ($T_\text{C}$), an abrupt superconducting phase transition ($\Delta T_\text{C}$) and a large critical current density $J_\text{C}$ \cite{Hadfield09}. Building upon recent studies \cite{Marsili09} we explored and optimised the deposition of NbN on GaAs using reactive DC magnetron sputtering and analysed the influence of growth conditions such as the growth temperature ($T_\text{growth}$) and N$_2$:Nb stoichometry on the crystal phase of thin films with thicknesses in the range $d_\text{NbN}=4-22$nm. By comparing our electrical measurements with detailed materials analysis using Rutherford Back Scattering (RBS) and X-Ray photoemission spectroscopy (XPS) we clearly show that the optimised films are both homogeneous and consist primarily of $\delta$-phase NbN \cite{Treece95}. We then continue to present the opto-electrical characterisation of SSPDs fabricated from \textit{optimised} films. By investigating the time-evolution of photon induced voltage pulses we deduce the hotspot resistance and the spatial extension of the hotspot along the nanowires. The results show that the devices fabricated from the $22$nm thick films exhibit a poor top-illumination detection efficiency ($0.004 \%$ at 950nm) but a low dark count rate ($\leq 3$cps). In contrast, detectors fabricated from much thinner $4$nm films exhibit detection efficiencies up to $21\pm2\%$, albeit with rather high dark count rates ($>50$kcps); values that compare very well to the current state of the art \cite{Gaggero10,Sprengers11} in this material system. The high detection efficiencies in combination with measured low timing jitters of $35$ps \cite{Zhang03} and the possibility for waveguide integration \cite{Sprengers11} makes these devices the ideal candidates for on-chip semiconductor based quantum photonics.

The NbN thin films investigated were grown using DC reactive magnetron sputtering in an Ar + N$_2$ plasma. Prior to growth, the chamber was conditioned by sputtering NbN at the desired nitrogen flux to remove contaminants and enhance the growth reproducibility. Before depositing the film, the GaAs [100] substrate was cleaned using acetone and isopropanol and baked for $4-6$h at $200^\circ{}$C to remove oxygen whilst remaining below the dissociation temperature of GaAs which occurs for temperatures of $\sim400^\circ{}$C \cite{Iizuka95}. During baking, the chamber was evacuated to a base pressure of $P_\text{base}\leq (2.0\pm1.0) \times 10^{-6}$mbar. After growth, the NbN film was electrically contacted in a 4-point geometry and the temperature dependence of the film resistance was recorded. 

%
\begin{figure}[t!]
\includegraphics[width=0.75\columnwidth]{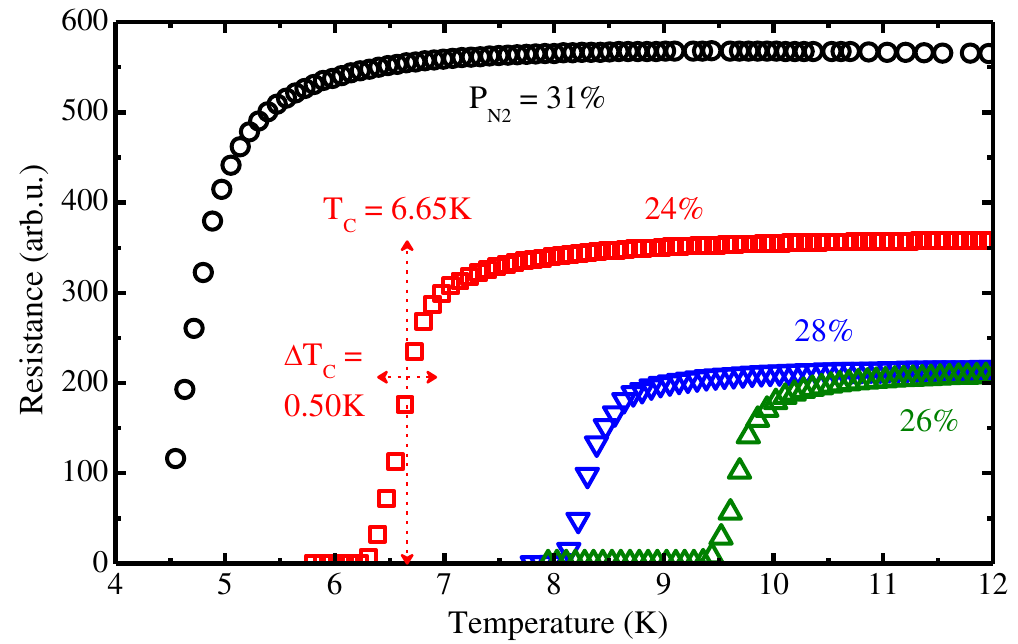}
\caption{\label{fig:Figure_1}
Temperature dependent NbN thin film resistance for nitrogen partial pressures in the range $26\%-31\%$ for films grown at a temperature of $T_\text{growth}=350^\circ $C with a thickness of $22$nm. The critical temperature $T_\text{C}$ and transition width $\Delta T_\text{C}$ are extracted using the mean value and width of the $90\% - 10\%$ resistance-transition width.}
\end{figure}
%

Figure \ref{fig:Figure_1} compares the temperature dependent resistance measured for a number of NbN films for a range of different nitrogen partial pressures $P_\text{N2}$ , determined in terms of the N$_2$ percentage of the total Ar + N$_2$ pressure ($P_\text{total} = 0.005\pm0.001$mbar). All films presented in fig \ref{fig:Figure_1} exhibit a thickness of $d_\text{NbN}=22\pm0.5$nm, determined by AFM, and were deposited at $T_\text{growth}=350^\circ{}$C. A clear superconducting phase transition is observed from which $T_\text{C}$ and $\Delta T_\text{C}$ can be extracted \footnote{The determination of $T_\text{C}$ and $\Delta T_\text{C}$ was done using the middle value and width of the $10\% - 90\%$ transition of the film resistance corresponding to \cite{MarsiliPhD09}.}. As $P_\text{N2}$ is reduced from $31\%$ to $26\%$, we observe a clear and systematic increase of $T_\text{C}$ from $4.8\pm0.2$K to $9.7\pm0.2$K and a commensurate decrease of $\Delta T_\text{C}$ ($0.63\pm0.03$K to $0.46\pm0.03$K). These observations indicate that the crystal quality improves as the N$_2$ partial pressure decreases. As $P_\text{N2}$ is reduced further from $26\%$ to $24\%$, $T_\text{C}$ again reduces to $6.7$K and $\Delta T_\text{C}$  increases, signifying a clear optimum nitrogen partial pressure at this growth temperature. The highest $T_\text{C}$ and lowest $\Delta T_\text{C}$ observed at $P_\text{N2}=26\%$ can be understood as the $P_\text{N2}$ that results in optimum film stoichiometry and microstructure for these growth conditions. A higher $P_\text{N2}$ pushes the ratio of N and Nb adatoms incorporated into the growth surface above the ideal stoichiometry and, thus, leads to a defective niobium sub-lattice. Similarly, lower $P_\text{N2}$ leads to a defective nitrogen sub-lattice \cite{Brauer61}. Since crystalline NbN has a critical temperature of $16.4$K \cite{Treece95} and thin ($5.5$nm) films on GaAs have been reported to exhibit $T_\text{C}$ up to $10.7$K \cite{Marsili09}, the maximum observed $T_\text{C}$ of $9.7$K for a $26\%$ nitrogen partial pressure is unlikely to correspond exclusively to the high-$T_\text{C}$ $\delta$-phase \cite{Thornton77} of NbN. As is confirmed below using materials microscopy, the partial pressure of $P_\text{N2}=26\%$ seems to favour a NbN crystal phase with a correspondingly lower $T_\text{C}$, such as the nitrogen-rich $\epsilon$-phase \cite{Thornton77}. In order to change growth conditions in a way that favours the formation of the desired $\delta$-phase, the growth temperature and $P_\text{N2}$ were systematically varied for a range of $4-6$nm (marked as $5\pm1$nm) and $21-23$nm (marked as $22\pm1$nm) thick films to fully explore the parameter space. 

%
\begin{figure}[t!]
\includegraphics[width=0.90\columnwidth]{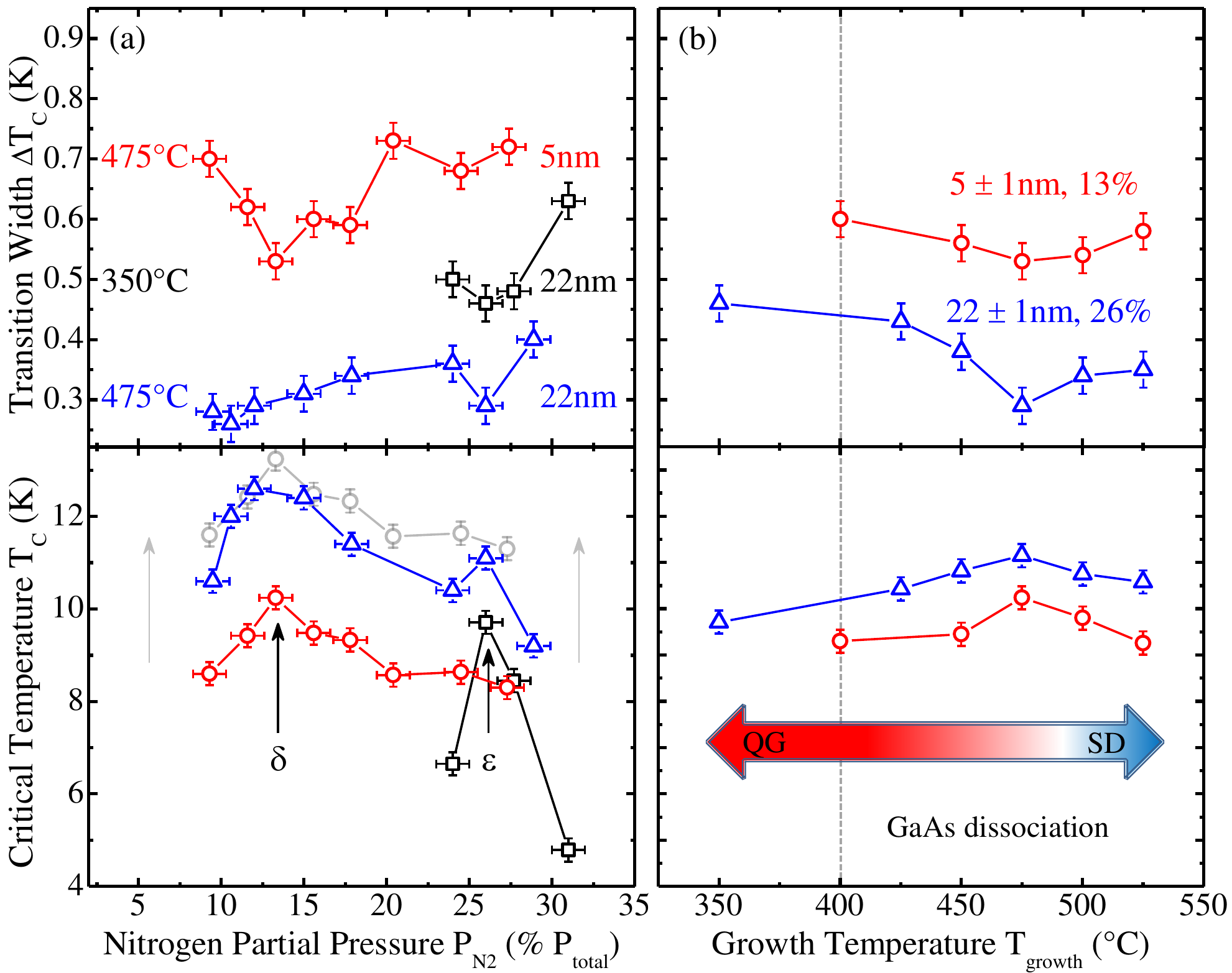}
\vspace{-15pt}
\caption{\label{fig:Figure_2}
Overview of measured superconductivity metrics for NbN films on GaAs with thicknesses of $4-6$nm (marked as $5\pm1$nm) and $21-23$nm (marked as $22\pm1$nm) as a function of $P_\text{N2}$ and $T_\text{growth}$. (a) Critical temperature $T_\text{C}$ (lower panels) and transition width $\Delta T_\text{C}$ (upper panels) measured for $5$nm (red symbols) and $22$nm (black/blue symbols) thick NbN films as a function of N$_2$ partial pressure $P_\text{N2}$ for $T_\text{growth}=350^\circ $C (black symbols) and $475^\circ $C (blue/red symbols). (b) Measured growth temperature dependent data.
}
\end{figure}
%
A summary of the results obtained during these $P_\text{N2}$ and $T_\text{growth}$ studies are presented in figs \ref{fig:Figure_2}(a) and \ref{fig:Figure_2}(b), respectively , where the lower panel shows the measured values of $T_\text{C}$ while the upper panel shows $\Delta T_\text{C}$. To test the reproducibility of the film deposition, we grew and measured three samples with a film deposited at $T_\text{growth}=475^\circ $C, $P_\text{N2}=26\%$. The results obtained were found to be highly reproducible, yielding a small error bar that was then adopted for all measurements presented in fig \ref{fig:Figure_2}. The $P_\text{N2}$ dependent measurements were performed for the thick films ($d_\text{NbN}=22$nm) at growth temperatures of $350^\circ $C (black squares) and $475^\circ $C (blue triangles), respectively. For the thin, $d_\text{NbN}=5$nm films, the nitrogen partial pressure was varied over the same range for substrate temperatures of $T_\text{growth}=475^\circ $C (red circles) only.  As already discussed in connection with fig \ref{fig:Figure_1}, the maximum $T_\text{C}$ measured for $P_\text{N2}=26\%$ and $350^\circ $C was $9.7$K, indicated by the arrow labelled $\epsilon$ in fig \ref{fig:Figure_2}(a).  Furthermore, for both thick and thin films grown at $475^\circ$C, a broad maximum is observed in $T_\text{C}$ at $P_\text{N2}=13\pm1\% $, indicated by the arrow labelled $\delta$ on fig \ref{fig:Figure_2}(a). A hint of the $\epsilon$ maximum is also observed at $P_\text{N2}\sim26\%$ for the thin films (fig \ref{fig:Figure_2}(a)). The observation of two maximum $T_\text{C}$ values for the datasets recorded with $P_\text{N2}=26\%$ and $13\%$, the prominence of which can be varied by changing $T_\text{growth}$, suggests that a change in film microstructure occurs from a nitrogen-rich phase at $P_\text{N2}=26\% $ to a phase with lower nitrogen content at $P_\text{N2}=13\% $.
Amongst all the different crystal phases that occur for NbN the face-centered cubic $\delta$-NbN$_x$ phase with a nitrogen content $0.88 \leq x \leq 0.98$ has the highest $T_\text{C}$ of $16.4$K \cite{Treece95}. Our observations suggest that the optimum $13\%$ nitrogen film grown at $T_\text{growth}=475^\circ$C may consist primarily of polycrystalline $\delta$-phase NbN, an expectation that is supported by the structural microscopy data presented below. The phase diagram of NbN$_x$ \cite{Brauer61} indicates that the hexagonal $\epsilon$-phase is preferentially formed for nitrogen contents, that are significantly higher than the $\delta$-phase \cite{Politis78}, lending support to the indentification of the two maxima labelled in fig \ref{fig:Figure_2}(a). Therefore, we tentatively attribute the two maxima in $T_\text{C}$ at $P_\text{N2}=26\%$, indicated by labelled arrows in fig \ref{fig:Figure_1}, as arising from $\delta$ and $\epsilon$-phases of NbN, respectively.

In order to obtain quantitative insight into the role of the \textit{growth temperature} in this picture, we systematically investigated the properties of $5$nm and $22$nm thick NbN-GaAs films grown at $P_\text{N2}=13\%$ and $26\%$ for growth temperatures in the range $350^\circ\text{C} \leq T_\text{growth} \leq 525^\circ$C. The results of these studies are summarised in fig \ref{fig:Figure_2}(b).  The maximum (minimum) $T_\text{C}$ ($\Delta T_\text{C}$) is found to occur close to $T_\text{growth}\sim475^\circ$C for all samples studied, independent of film thickness or the nitrogen partial pressure. This global maximum can be interpreted as reflecting the interplay of two effects; Firstly, for $T_\text{growth}\geq 400^\circ $C desorption of As$_2$ from the GaAs $[100]$ surface begins \cite{Bayliss76} promoting surface reconstruction and a deviation from ideal stoichiometry - both effects that may help to overcome the large lattice mismatch between NbN and GaAs. Secondly, the diffusion length of adatoms arriving at the growth surface significantly increases with $T_\text{growth}$ promoting a transition from the quenched growth regime, in which poor crystal quality arises from the incorporation of adatoms on the surface close to point where they arrive, to a more favourable growth regime dominated by surface diffusion \cite{Thornton77,MarsiliPhD09}. In the quenched growth regime, the microstructure typically consists of columns of material with poor crystallinity, and even amorphous regions separated by voids while higher $T_\text{growth}$ significantly promotes improved crystal quality with isotropic equiaxed crystal grains.  Generically, NbN films grown at much higher temperatures ($\geq600^\circ $C) on conventional substrates such as MgO and sapphire tend to exhibit high $T_\text{C}$ \cite{Chockalingam08} as a consequence of the dominance of the surface diffusion limited growth regime and resulting high crystal quality. The observed maximum (minimum) of $T_{C}$ ($\Delta T_\text{C}$) behaviour is found to be independent of the nitrogen partial pressure during growth, as shown by the similarity of the blue ($P_\text{N2}=26\%$, $d_\text{NbN}=22$nm) and red ($P_\text{N2}=13\%$, $d_\text{NbN}=5$nm) curves in fig \ref{fig:Figure_2}(b).  


\begin{figure}[t!]
\includegraphics[width=0.60\columnwidth]{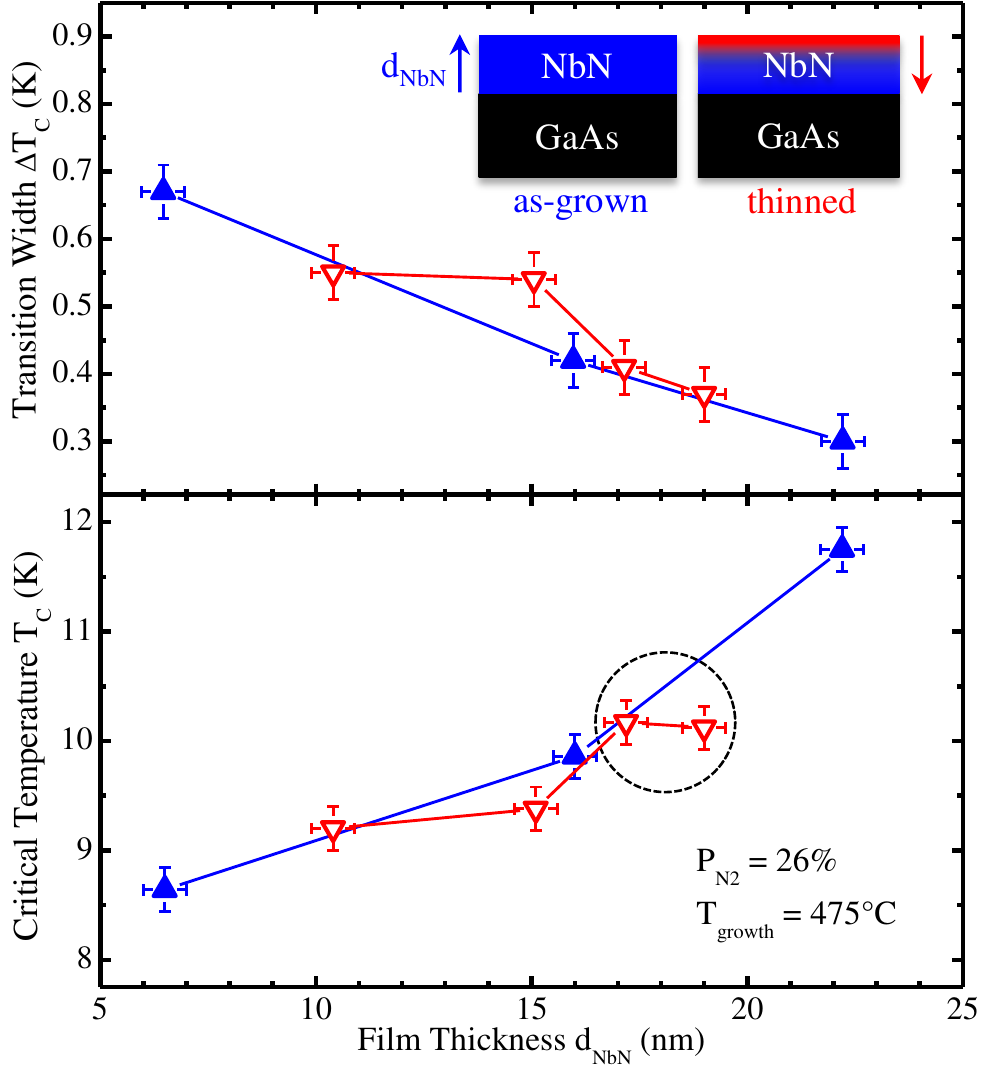}
\caption{\label{fig:Figure_3}
Critical temperature $T_\text{C}$ (bottom panel) and transition width $\Delta T_\text{C}$ (top panel) as a function of film thickness $d_\text{NbN}$ for NbN films grown at $P_\text{N2}=26 \% $ and $T_\text{growth}=475^\circ $C. Blue triangles correspond to as-grown films with sputtering times of $15\text{s}, 30\text{s}$ and $55\text{s}$, indicated in the schematic on the top. The red symbols correspond to selectively thinned films, originating from a single $19$nm film, which was etched down for $0\text{s}, 30\text{s}, 60\text{s}$ and $90\text{s}$, respectively.}
\end{figure}


To support these viewpoints and exclude the possibility that the crystal structure and quality vary strongly with $d_\text{NbN}$ due to the formation of a buffer layer, comparative measurements were performed on several films grown under nominally identical conditions, but with different final thicknesses ($6$nm $\leq d_\text{NbN}\leq22$nm), and a $19$nm thick film that was \textit{selectively thinned} after growth using reactive ion etching in a $\text{SF}_6$/$\text{C}_4\text{F}_8$ plasma.  In the following we refer to these two types of samples as the "as-grown" and "thinned" films, respectively. Figure \ref{fig:Figure_3} shows $T_\text{C}$ (lower panel) and $\Delta T_\text{C}$ (upper panel) as a function of film thickness for as-grown (blue triangles) and thinned (red triangles) films. All the NbN films studied in fig \ref{fig:Figure_3} were grown using $P_\text{N2}=26\% $ and $T_\text{growth}=475^\circ{}$C. The as-grown films had a thickness $d_\text{NbN}=22\text{nm}, 16\text{nm}$ and $6$nm, whereas the thinned films were obtained from an initially $19$nm thick film that was etched back for $0\text{s}, 30\text{s}, 60\text{s}$ and $90\text{s}$ resulting in film thicknesses of $19$nm, $17$nm, $15$nm and $10$nm, as measured by AFM. Comparing $T_\text{C}$ and $\Delta T_\text{C}$ of as-grown and thinned films reveals results that are identical within the experimental error. For as-grown films $T_\text{C}$ decreases from $11.7$K to $8.6$K as $d_\text{NbN}$ reduces from $\sim22$nm to $\sim6$nm. Over the same thickness range $\Delta T_\text{C}$ increases from $0.30$K to $0.67$K. For thinned films $T_\text{C}$ decreases from $10.1$K to $9.2$K, whereas $\Delta T_\text{C}$ increases from $0.37$K to $0.55$K. These trends are in very good agreement with the findings of \cite{Marsili09}, which showed that the reduction of $T_\text{C}$ (increase of $\Delta T_\text{C}$) reflects the stronger influence of disorder which induces the superconductor-insulator transition \cite{Larkin99, Markovic99}.

The observation of a monotonic $d_\text{NbN}$ dependence of the superconducting metrics, that are identical for as-grown and thinned films, indicates that the grown films are homogeneous along the growth direction. In addition, the thickness of the NbN film appears to have little or no influence on the growth conditions indicating that the large lattice mismatch between NbN and GaAs is most likely accomodated by defects close to the NbN-GaAs interface. In contrast, crystal quality seems to be intrinsically limited for thinner films.  The finding that the film thickness has no significant influence on the growth conditions of NbN on GaAs is not at all intuitive, since one would expect a thickness dependence caused by the strain field that builds up due to the $27\%$ lattice mismatch \cite{Marsili09} at the NbN-GaAs interface. 
Considering these findings and the data presented in fig \ref{fig:Figure_2}, we conclude that the nitrogen partial pressure fully determines the crystal phase, whereas a growth temperature of $\sim475^\circ$C provides an optimum crystal quality independent of film thickness.

The clear film thickness dependence of the superconducting metrics presented in fig \ref{fig:Figure_3} allows us to quantitatively scale the data for the $5$nm thick films presented in fig \ref{fig:Figure_2} and directly compare it with the data recorded from the $22$nm film. The result of this procedure is shown by the grey curve plotted in fig \ref{fig:Figure_2}(a).  Remarkably, both the form of the data recorded from the scaled $5$nm data and the absolute values of $T_\text{C}$ ($\Delta T_\text{C}$) agree within the experimental errors strongly suggesting that the observed maximum in the film quality for $P_\text{N2}\sim13\%$ is indeed a general result.

\begin{figure}[t!]
\includegraphics[width=0.90\columnwidth]{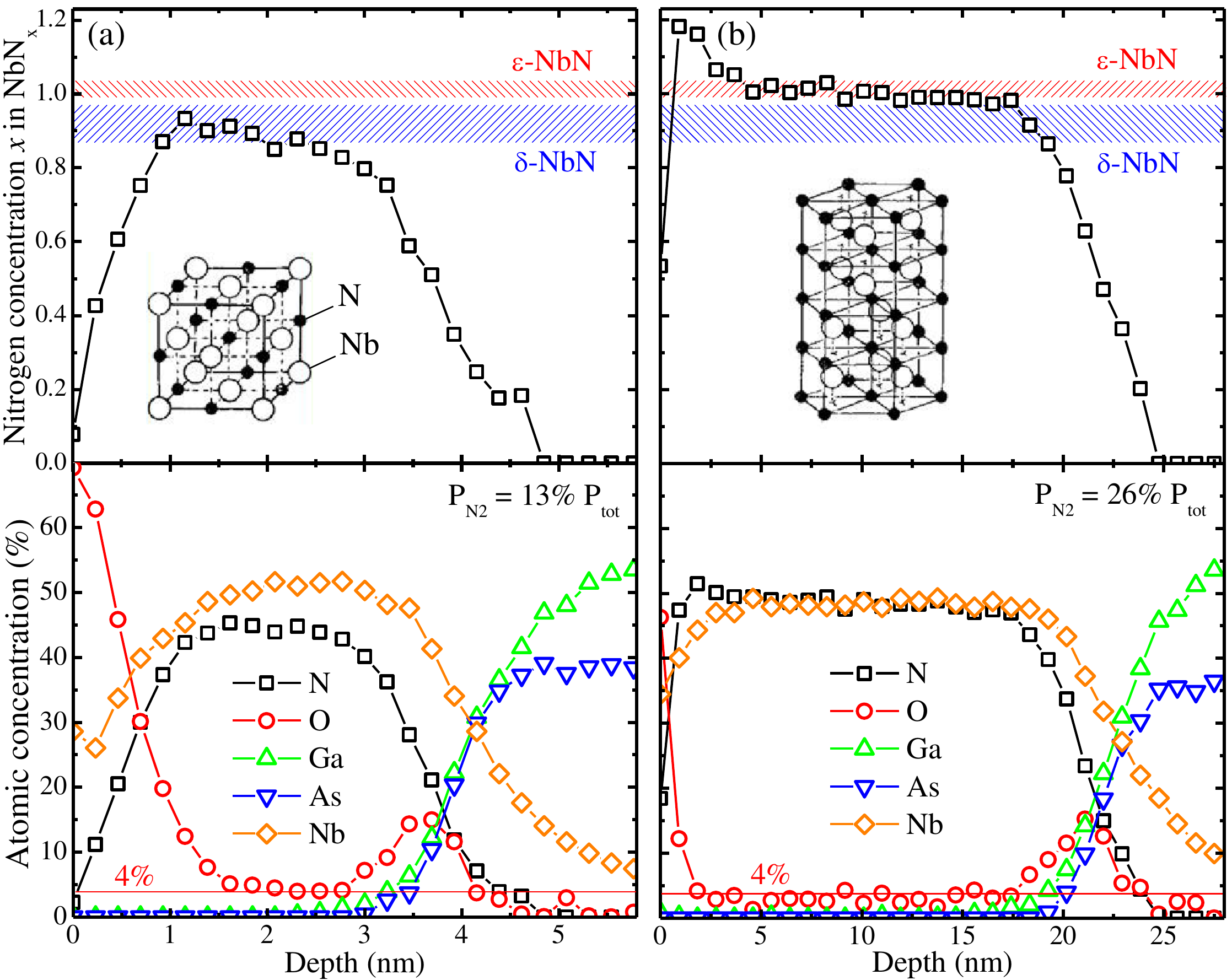}
\caption{\label{fig:Figure_4}
Atomic concentration (bottom panels) and nitrogen content $x$ (top panels) for NbN$_x$ films as a function of depth obtained by XPS and RBS. (a) corresponds to $d_\text{NbN}=4$nm grown at $P_\text{N2}=13 \% $ and (b) shows $d_\text{NbN}=22$nm grown at $P_\text{N2}=26 \% $. Both films were grown for $T_\text{growth}=475^\circ $C. $P_\text{N2}=13 \% $ grown films consist of nitrogen poor fcc $\delta$-NbN, as indicated by the blue shaded region, whereas $P_\text{N2}=26 \% $ grown films exhibit nitrogen rich hexagonal $\epsilon$-NbN, as indicated by the red shaded region. The corresponding crystal structures are adapted from \cite{Brauer61}.
}
\end{figure}%

We continue by presenting direct measurements of the depth profile of the atomic composition and stoichiometry of the NbN films grown using the optimum growth temperature of $T_\text{growth}=475^\circ$C and obtain results that strongly support these conclusions.  Depth dependent X-ray photoemission spectroscopy (XPS) and Rutherford backscattering (RBS) measurements were performed through the NbN film, and into the GaAs substrate, to confirm that the nitrogen concentration during growth does indeed determine the crystal phases and that the optimised films consist primarily of $\delta-$phase NbN.  Measurements were performed on two films grown with $13\%$ ($4$nm thick) and $26\%$ ($22$nm thick) nitrogen partial pressures as marked by arrows on fig \ref{fig:Figure_2}(a). Figures \ref{fig:Figure_4}(a) and \ref{fig:Figure_4}(b) show the depth dependent atomic concentrations (bottom panels) and nitrogen content $x$ (top panels) for the $13\%$ ($4$nm thick) and $26\%$ ($22$nm thick) films, respectively.
The XPS measurements reveal a thin NbO$_x$ layer at the surface with a thickness of $\sim1$nm. This finding compares very well with the observations from the etching depth dependent measurements presented in fig \ref{fig:Figure_3} that showed no difference between the unetched sample and the sample etched for $30$s, as indicated by the dashed circle.
While the oxygen concentration decreases rapidly with depth, to levels below $4\%$ for the thick film, the oxygen concentration remains above $\sim4\%$ for the thin film throughout the NbN layer. This can be clearly seen by the red curve in fig \ref{fig:Figure_4}. Since both samples were grown for the same base pressure of $P_\text{base}<3\cdot 10^{-6}$mbar, the oxygen concentration should be equal for both films and, hence,  we interpret the higher oxygen concentration in the thin film to post-growth oxygen interdiffusion. Whilst the oxygen concentration decreases, the Nb and N content in both samples increases and then remains constant throughout the entire film.
At the NbN-GaAs interface we again observe a thin oxygen rich layer with a thickness of $\sim1$nm and the stoichiometry of the GaAs layer deviates from the expected value of $50\%$. This observation correlates very well with the thermal dissociation of GaAs at the surface for $T_\text{growth}\geq 400^\circ$C due to As$_2$ desorption. The resulting arsenic vacancies appear to be occupied by Nb atoms, as shown in the atomic concentrations in the bottom panels of fig \ref{fig:Figure_4}. 
The top panels of fig \ref{fig:Figure_4} show the nitrogen concentration $x$ of the NbN$_x$ films as a function of depth for the film grown at $13\%$ N$_2$ (left panel) and $26\%$ N$_2$ (right panel). The range of $x$ that would correspond to $\delta$-NbN \cite{Brauer61,Politis78} is indicated schematically by the blue hatched region, whereas the range corresponding to the $\epsilon$-phase \cite{Brauer61,Politis78} is denoted by the red shaded region, respectively. For the thin film, grown at $13\%$ N$_2$, $x$ increases from $0$ to $\sim0.89$ within the $1$nm thick region of the film closest to the surface. Thereafter, $x$ slowly decreases again for depths ranging from $2$ to $3$nm but remains entirely within the range expected for $\delta$-NbN. Afterwards, a sharp decrease in $x$ is observed close to the NbN-GaAs interface. From these measurements we conclude that the film grown for $P_\text{N2}=13\%$ does indeed consist of the $\delta$-phase having high $T_\text{C}$, in good agreement with observations from the electrical transport measurements and the optimised superconductivity metrics presented in figures \ref{fig:Figure_1} and \ref{fig:Figure_1}. The data for the sample grown at $P_\text{N2}=26\%$, presented in the rightmost column of fig \ref{fig:Figure_4}, exhibits a similarly sharp increase in $x$ over the uppermost $1$nm. However, in contrast to the thin sample a much higher value of $x\sim1.2$ is reached. Afterwards, $x$ gradually decreases to $\sim1.05$ within the next $\sim 2$nm and then remains constant in the range expected for the $\epsilon$-phase, until the NbN-GaAs interface is reached. These observations support the conclusion that films grown with $P_\text{N2}=26\%$ consist primarily of the nitrogen-rich $\epsilon$-phase, supporting the conclusions reached in relation to the discussion of fig \ref{fig:Figure_2}.


\begin{figure}[t!]
\includegraphics[width=0.85\columnwidth]{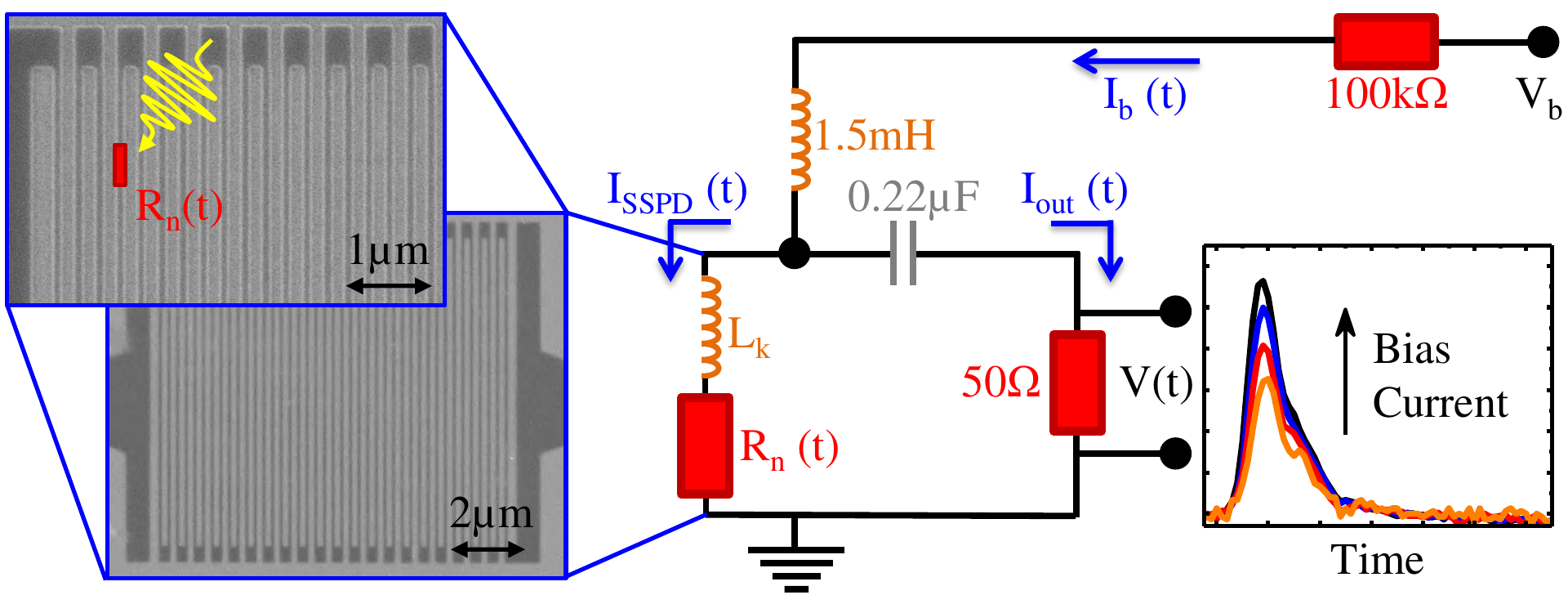}
\caption{\label{fig:Figure_5}
Schematic of opto-electrical SSPD characterisation setup. A meander-type detector (SEM pictures on the left) is operated just below the critical current $I_\text{C}$ by using a bias-tee ($1.5$mH inductance and $0.22\text{$\micro$F}$ capacitance) connected to a $100$k$\Omega$ resistor and a constant voltage source $V_\text{b}$. The SSPD can be modelled as an inductor $L_\text{k}$ and a time-dependent resisitor $R_\text{n}(t)$. As depicted schematically in the leftmost SEM image, a normal conducting region $R_\text{n}$ is formed upon photon absorption. This leads to a redirection of the bias current onto the output-capacitor, producing a voltage pulse $V(t)$ which is monitored using an oscilloscope.
}
\end{figure}


Using optimised, predominant $\delta$-phase NbN films we fabricated SSPDs in the typical meander-type geometry \cite{Goltsman01} on $4$nm and $22$nm thick films. In the following, these two samples are termed the thin and thick SSPDs, respectively. The SSPD fabrication was performed using a negative tone electron beam resist for patterning of the detectors. After development of the exposed resist, the sample is reactively etched in a $\text{SF}_6$/$\text{C}_4\text{F}_8$ plasma, transferring the electron beam written structures onto the NbN film. This results in $10\times10\text{$\micro$m}^2$ NbN meanders on GaAs, each of which consists of $41$ NbN nanowires with a width of $100\pm5$nm, as presented in the SEM images in fig \ref{fig:Figure_5}. In order to electrically address single detectors in a low-temperature microwave probe station, Ti/Au contact pads were defined using photo-lithography.
For opto-electrical characterisation, the sample was cooled down to $T=4.3\pm0.1$K and single detectors are electrically connected to the bias-tee that has an inductance of $1.5$mH and a capacitance of $0.22\text{$\micro$F}$ (fig \ref{fig:Figure_5}). A fixed bias-voltage $V_\text{b}$ in series with a $100$k$\Omega$ resistor is fed into the DC port of the bias-tee, operating the detector at a drive-current  $I_\text{b}=V_\text{b}/100\text{k}\Omega$. The AC port of the bias-tee is directly connected to a $20$GHz sampling oscilloscope with a $50\Omega$ input impedance for pulse-shape measurements, as depicted schematically in fig \ref{fig:Figure_5}. Such meander type nanowire detectors can be electrically modelled as an inductor $L_\text{k}$ in series with a time-dependent resistor $R_\text{n}(t)$ \cite{Kerman06,Yang07}. When operating the device with a bias current $I_\text{SSPD}=I_\text{b}$, slightly below the device-critical current $I_\text{C}$, the absorption of a photon by a Cooper pair creates a small normal conducting region, a so-called hotspot, which then extends across the whole wire \cite{Skocpol74,Semenov01}. This process is schematically depicted in the SEM picture in the upper left panel of fig \ref{fig:Figure_5}. The appearance of $R_\text{n}\neq0$ in the left arm of the circuit redirects a part of the bias current $I_\text{out}=I_\text{b}-I_\text{SSPD}$ into the right arm leading to a voltage pulse $V(t)$ at the oscilloscope. Due to the ohmic character of $R_\text{n}$, higher bias currents result in higher voltage pulses, as shown in the $V(t)$ data presented in fig \ref{fig:Figure_5}. After hotspot healing, the recovery of the current is limited to a nanosecond timescale by the rise time $\tau_\text{I,rise}=L_\text{k}/50\Omega$ \cite{Kerman06}, thereby limiting the detector countrate to a few hundred MHz.



\begin{figure}[t!]
\includegraphics[width=0.9\columnwidth]{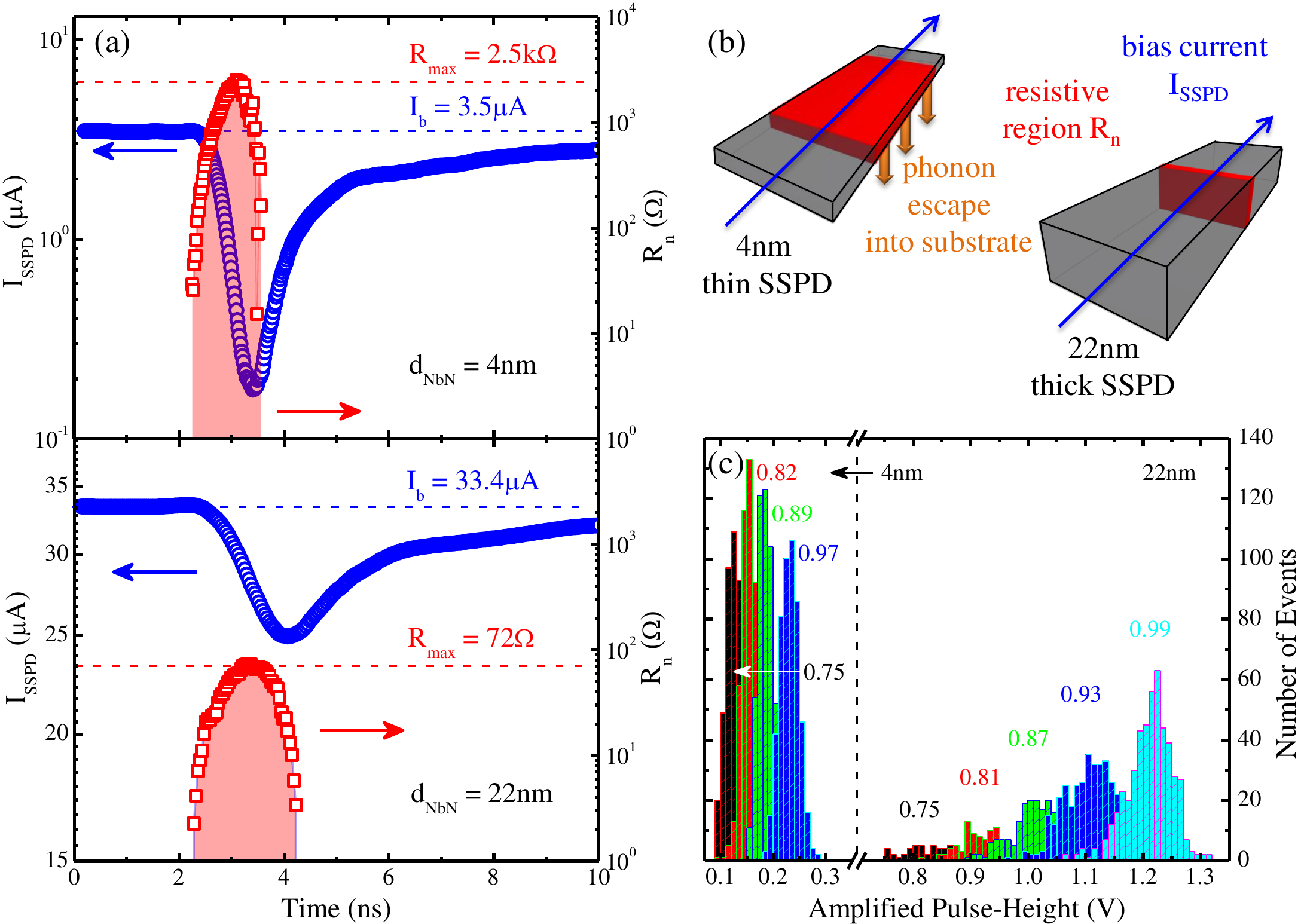}
\caption{\label{fig:Figure_6}
(a) Temporal dependence of the hotspot resistance (red squares) and detector bias current (blue circles) for the thick (bottom panel) and thin (top panel) SSPD. (b) Schematic of a photon induced resistive region along the nanowire. (c) Histograms of the amplified voltage pulses resulting from hotspot detection events for different bias currents $I_\text{b}$ in units of $I_\text{C}$. The leftmost (rightmost) data sets were obtained from SSPDs defined with $4\pm0.5$nm/$22\pm0.5$nm thick films.
}
\end{figure}

To probe the time evolution of the hotspot resistance, both thick and thin SSPDs were operated and measured using the circuit presented in fig \ref{fig:Figure_5}. Laser pulses with a duration of $\sim90$ps at $\lambda=950$nm were delivered by a laser diode with a repetition rate of $80$MHz. Light is focussed onto the detector using a long working distance compound objective. 
Typical voltage pulses $V(t)$ arising from the response of the thick and thin detectors are shown in the inset of fig \ref{fig:Figure_5}.
For the thick detector the observed pulses could be detected directly by a $20$GHz sampling oscilloscope, whereas voltage traces originating from the thin SSPD were amplified using two calibrated low-noise $30$dB high-frequency amplifiers.
For the further analysis, the amplification and pulse shortening due to high-pass filtering of the amplifiers were corrected to reproduce the original, photo-generated pulse shape.
The kinetic inductance of the corresponding device was determined by fitting an exponential decay to the falling edge of the voltage trace.
The thick SSPD then exhibits $L_\text{k}=76\pm14$nH, whereas the thin SSPD shows $L_\text{k}=393\pm20$nH, the difference between the two values being in very good agreement with $L_\text{k}=\mathcal{L}_\text{k}\int ds / A(s)$, where $\mathcal{L}_\text{k}$ is the kinetic inductivity, $A$ is the cross-sectional area and integration is done along the wire \cite{Kerman06,Hadfield05}.
The temporal evolution of the hostpot resistivity $R_\text{n}(t)$ and the corresponding current drop $I_\text{SSPD}(t)$ can then be obtained from $V(t)$ by using a simple electro-dynamic model \footnote{
Following the electrodynamic model presented in \cite{Yang07}, $R_\text{n}(t)$ and $I_\text{SSPD}(t)$ can be derived by using the measured voltage trace $V(t)$ to solve the following equations, obtained from the circuit presented in fig \ref{fig:Figure_5}:\\ 
$I_\text{b}(t) = I_\text{SSPD}(t) + I_\text{out}(t) \text{ with } I_\text{out}(t) = V(t) / 50 \Omega$\\
$V_\text{b}  = V_\text{high}(t) + V_\text{low}(t) $\\
$V_\text{high}(t) = 100\text{k}\Omega \times I_\text{b}(t) + 1.5\text{mH} \times \dot{I}_\text{b}(t)  \approx 100\text{k}\Omega \times I_\text{b}$\\
$\dot{V}_\text{low}(t) = 1/0.22\text{$\micro$F} \times I_\text{out}(t) + \dot{V}(t)$ \\
$\Rightarrow \ I_\text{SSPD}(t) = (V_\text{b}-V_\text{low}(t))/100\text{k}\Omega - V(t)/50\Omega $\\
$\Rightarrow \ R_\text{n}(t) = 1/I_\text{SSPD}(t) \left (  V_\text{low}(t) - L_\text{k} \times \dot{I}_\text{SSPD} \right )$
}
for the circuit shown in fig \ref{fig:Figure_5}.
The data was recorded for a fixed bias current of $I_\text{b} = 0.56 I_\text{C}$ for the thin SSPD and $0.75 I_\text{C}$ for the thick one, as indicated by the dashed blue lines in fig \ref{fig:Figure_6}a.
The current through the detector $I_\text{SSPD}(t)$ (blue circles) and the hotspot resistance $R_\text{n}(t)$ (red squares) are presented in a logarithmic scale in fig \ref{fig:Figure_6}a for the thin detector (top panel) and the thick detector (bottom panel), respectively.
For the thick detector the resistance builds up within $1.3$ns to a maximum value of $R_\text{max}=72\Omega$ and then reduces to zero within $\sim0.8$ns.
In contrast, the thin SSPD shows a $35\times$ larger maximum resistance of $2.5\text{k}\Omega$ which is $\sim 50\%$ smaller than typical values for SSPDs on MgO substrates \cite{Yang07}.
Furthermore, when compared to the thick device the resistance build-up and healing takes place over much faster timescales for the thin devices ($0.7$ns and $0.4$ns, respectively).
For both cases, a drop of $I_\text{SSPD}$ is observed shortly after the hotspot resistivity starts to build up, delayed by the kinetic inductance of the device. 
For the thick detector, the current drops to $74\%$ ($34 - 25\text{$\micro$A}$), whereas the current through the thin detector reduces to $6\%$ ($3.4 - 0.2\text{$\micro$F}$), reflecting the much higher resistance of the thin device.
Again, delayed by the kinetic inductance, the current then recovers with a rise time of $1.5\pm0.3$ns for the thick detector and $7.9\pm0.4$ns for the thin one, the difference arising from the film thicknesses.
However, the $34\times$ larger hotspot resistivity of the thinner detector cannot be explained by a $5.5\times$ smaller film thickness alone.
As the build-up of a resistive region across the entire wire upon photon absorption is determined by a complex interplay of various effects \cite{Goltsman05,Yang07,Goltsman01}, we present here a qualitative explanation for the observed behaviour.
To get a better understanding of the responsible mechanisms, the horizontal extension of the hotspot along a nanowire is estimated for both cases.
With the knowledge of the ohmic resistance $R_\text{wire}$ of a single nanowire of length $l_\text{wire}$ for temperatures $T>T_\text{C}$, the horizontal extension of the photon induced resistive region $l_\text{max}$ along the nanowire can be estimated to be $l_\text{max}=l_\text{wire} \times R_\text{max}/R_\text{wire}$.
Here, $R_\text{wire}$ is determined from the current voltage characteristics of the investigated devices where a sequential switching of the nanowires from superconducting state to normal conductivity is observed.
Now $l_\text{max}$ can be estimated to be $270\pm10$nm for the thin detector and $30\pm8$nm for the thick detector, as schematically shown in fig \ref{fig:Figure_6}b.
Here, the larger hotspot extension of the thinner detector promotes a faster cooling of the resistive region via phonon emission into the adjacent GaAs substrate.
In addition, thermal boundary conductivity at the NbN/GaAs interface scales with $1/d_\text{NbN}$ \cite{Yang07} and, hence, is expected to be significant higher for the thin device.
In case of the thick detector the initially photon-created hotspot with an extension of $\sim13$nm \cite{Maingault10} cannot extend as fast and wide as for the thin device due to heat transfer within the wire \cite{Yang07}. Here, cooling inside the comparatively larger heat sink provided by the surrounding nanowire takes place, before switching of the side channels \cite{Goltsman05,Goltsman01} and creation of the entire resistive region occours.
Therefore, if the photon absorption leads to a detectable event, the resistance and its lateral extension are expected to be smaller, since hotspot quenching is promoted during the longer resistance built up, as observed in fig \ref{fig:Figure_6}a.
These considerations suggest a considerably lower dark count rate for thick SSPD devices, however, at the price of a much smaller detection efficiency.

\begin{figure}[t!]
\includegraphics[width=0.90\columnwidth]{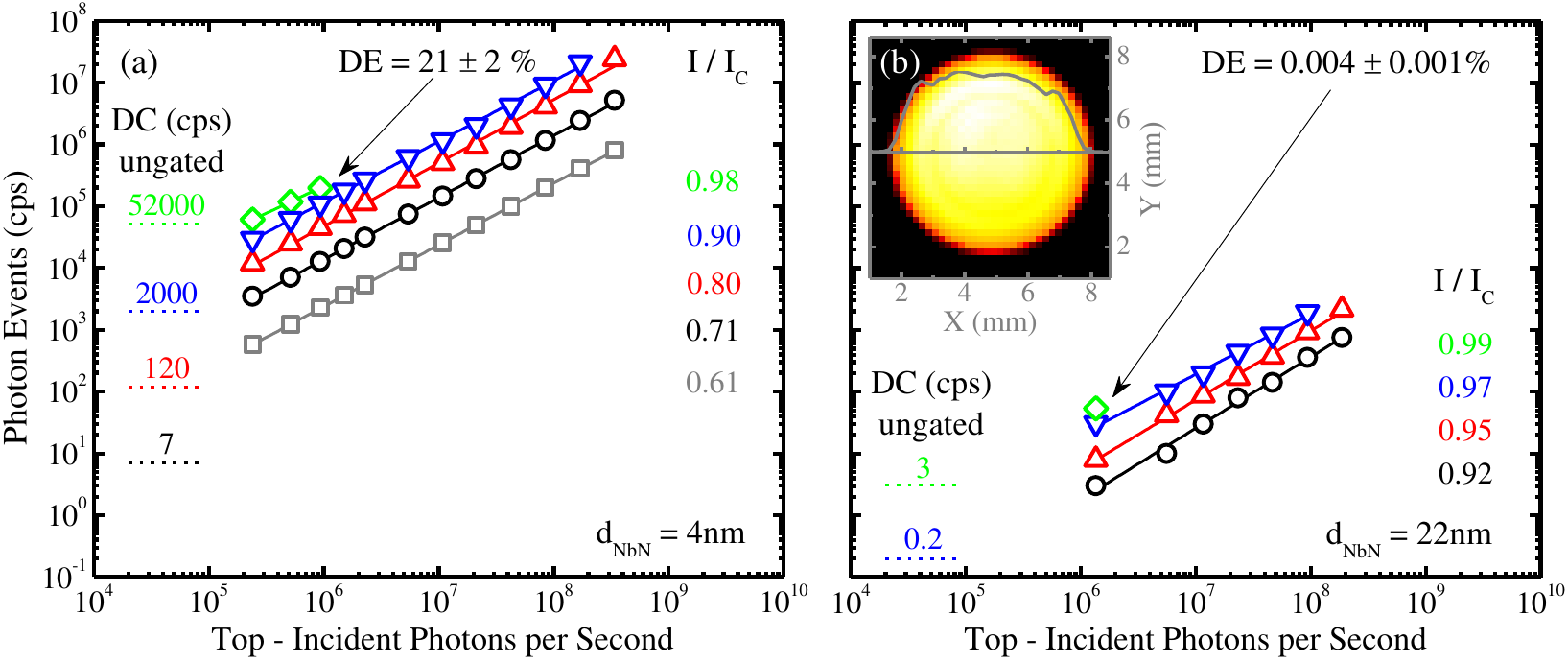}
\caption{\label{fig:Figure_7}
Number of detected photon events as a function of the top-incident photon number for a $4$nm  thin SSPD (a) and a $22$nm thick SSPD (b), respectively. Data is shown in a double logarithmic plot and presented for different $I/I_\text{C}$ ratios. The corresponding dark count rates are indicated by the dashed lines on the left. The solid lines are linear fits with slopes of $0.98\pm0.02$ (thin device) and $1.09\pm0.03$ (thick device) revealing the single photon character of the detected events. A maximum detection efficiency of $21\pm2\%$ is reached for the $4$nm detector at $0.98 I_\text{C}$, whereas the $22$nm SSPD shows $0.004\pm0.001\%$ detectivity for a current of $0.99 I_\text{C}$. The inset shows the measured beam profile used for illumination.
}
\end{figure}

In order to determine the detection efficiency, we set an event-treshhold level $V_\text{t}$ that both rejects the noise of the readout-electronics but does not cut off any photon detection events.
In order to find the optimum value of $V_\text{t}$, the device was illuminated with a constant calibrated photon flux and a histogram of the amplified voltage pulses of the detection events was recorded for different values of $I/I_\text{C}$. 
The results of typical measurements are presented in figure \ref{fig:Figure_6}c for the thin (leftmost panel) and thick (rightmost panel) detectors.
For these measurements the maximum output voltage was recorded within each trigger cycle, as given by the laser repetition rate.
The data presented shows the average over 500 independent measurements. 
In this way, the number of events depicted in the histogram reflects the number of registered photon detection events per time and, therefore, when compared with the calibrated incident photon flux on the detector, the corresponding detection efficiency. 
As one would expect from other reports in the literature \cite{Divochiy08,Dorenbos11}, the detectivity scales exponentially with the drive current, an effect that can be clearly seen in the data recorded for the thick detector.
In contrast, the thin detector was operated in saturation, so the histograms show no increase of the registered events for higher drive currents.
Besides, a clear shift of the mean pulse height is observed for increasing bias currents reflecting the ohmic nature of the photon induced resistive region. For the following count-rate measurements, presented in figure \ref{fig:Figure_7}, the voltage treshhold level was set to be $80$mV for the thin SSPD and to $700$mV for the thick detector, respectively.


Before discussing further experimental findings of thick and thin detectors, we compare these devices with respect to their critical current densities $J_\text{C}=I_\text{C}/A$ (with the wire cross-section $A$) and find the ratio $J_\text{C}^\text{thick}/J_\text{C}^\text{thin}=1.3\pm0.2$. However, since $J_\text{C}$ is expected to be solely a property of the crystal phase, this ratio should be close to unity. By defining an \textit{effective active film thickness} $d_\text{eff}=d_\text{NbN}-d_\text{Ox}$, where $d_\text{eff}$ is the active thickness of the superconducting NbN layer and $d_\text{Ox}$ is an optically and electrically dead oxide layer, we found that this ratio becomes unity for $d_\text{Ox} \sim 1$nm, in excellent agreement with the XPS results presented in fig \ref{fig:Figure_4}. Therefore, SSPDs fabricated from the $4$nm film have an effective thickness $d_\text{eff}=3\pm0.5$nm and, due to their extremely small thickness, are expected to show very high detection efficiencies \cite{Hofherr10}.
For the detectivity and dark count measurements, presented in figure \ref{fig:Figure_7}, the oscilloscope in the circuit in fig \ref{fig:Figure_5} is replaced by two low-noise $30$dB high-bandwidth amplifiers which are connected to a $350$MHz frequency counter. 
The respective SSPD is now operated at a fixed bias current ratio $I/I_\text{C}$, as indicated for each data series presented in fig \ref{fig:Figure_7}.
The device is illuminated using a parallel laser beam with a carefully calibrated circular intensity profile with a diameter of $d_\text{beam}=5.5\pm0.1$mm (see fig \ref{fig:Figure_7}b - inset).
The recorded number of events per second (open symbols in fig \ref{fig:Figure_7}) is then corrected for the respective dark count rate (dashed line in fig \ref{fig:Figure_7}).
In order to calibrate the incident photon arrival rate on the detector, the beam profile was mapped by scanning the beam accross a $500\pm100\text{$\micro$m}$ diameter pinhole and measuring the transmitted power profile, as shown in the inset of fig \ref{fig:Figure_7}b.
A horizontal cross-section of the beam profile at $Y=5$mm is shown in red.
Now the number of incident photons on the detector can be calculated using $N_\gamma=P/(\hbar \omega) \times A_\text{SSPD}/(\pi d^2_\text{beam}/4) \times x_\text{loss} \times x_\text{beam}$ with the average power of the beam $P$, the photon energy $\hbar\omega$, the active area of the detector $A_\text{SSPD}$, the measured losses at cryostat windows $x_\text{loss}$ and the correction factor arising from the detector position within the beam profile $x_\text{beam}$.
For both the $4$nm thin (fig \ref{fig:Figure_7}a) and the $22$nm thick (fig \ref{fig:Figure_7}b) detector we observe a clear linear increase of photon detection events as a function of the incident photon number in this double logarithmic plot for all shown bias current ratios, proving that the detectors operate in the single photon regime \cite{Maingault10,Divochiy08}.
The solid lines are fits with an exponent of $0.98\pm0.02$ for the thin SSPD and $1.09\pm0.03$ for the thick one, respectively.
The fact that the observed behaviour is slightly superlinear for the thick device suggests that a small percentage of detected events arises from double photon events, in agreement with expectations for larger detector areas \cite{Maingault10}. 
For both the thin and thick SSPDs the dark count rate as well as photon counts at a fixed photon flux scale exponentially with $I/I_\text{C}$, which is in good agreement with previous publications \cite{Divochiy08,Dorenbos11}.
In order to estimate the maximum device detection efficiency, the number of detected events is divided by the number of incident photons at $0.98I/I_\text{C}$ for the thin, $4$nm detector and for $0.99I/I_\text{C}$ for the thick, $22$nm one, respectively.
This was done for the maximum possible photon flux that did not push the detector permanently into a normal conducting state caused by a reduction of the critical current due to device heating \cite{Yang07}.
In this way the detection efficiency $\eta$ for top-illumination using light at $950$nm was determined to be $\eta=21\pm2\%$ for the thin detector at an ungated dark count rate of $52000$ per second.
This compares very well to the current state-of-the-art NbN SSPD detectors on GaAs that have been reported to have a normal incidence detection efficiency of $18.3\%$ at $1300$nm \cite{Gaggero10}, as well as to findings reported for other substrates \cite{Kor05,Goltsman05,Goltsman01}.
Also the high dark count rate compares well with other values reported in literature \cite{Sprengers11}.
In contrast, the thick SSPD exhibits a much lower detection efficiency $\eta=0.004\pm0.001\%$ with an almost negligible dark count rate of only $3$ per second. This observation is fully expected from the hotspot analysis presented in figure \ref{fig:Figure_6} and is in very good accord with the findings reported in \cite{Hofherr10}.

In summary, we optimised the growth of NbN thin films on GaAs substrates  by varying the substrate temperature and $N_2$ partial pressure. 
Optimum crystal phases for SSPD operation were obtained using a nitrogen partial pressure of $P_\text{N2}=13\%$ at a growth temperature of $T_\text{growth}=475^\circ$C. Evidence was obtained that these conditions promote the formation of the high $T_\text{C}$ $\delta$-phase of NbN.
Under these growth conditions an optimised $d_\text{NbN} = 4\pm0.5$nm thin film with $T_\text{C}=10.2\pm0.2$K and $\Delta T_\text{C}=0.53\pm0.03$K has been obtained, values which are very close to the state of the art on GaAs substrates \cite{Marsili09}. For lower $T_\text{growth}$ the superconducting critical temperature reduces significantly. The maximum $T_\text{C}=9.7$K for $22$nm films for $P_\text{N2}=26\%$ and $T_\text{growth}=350^\circ{}$C was attributed to the formation of the hexagonal $\epsilon$-phase with significantly lower $T_\text{C}$ values, as compared to $\delta$-NbN. 
By systematically varying $T_\text{growth}$, we observe an optimum value of $475^\circ{}$C, independent of nitrogen partial pressure, attributed to a trade-off between disorder induced by arsenic evaporation and higher crystal quality caused by increased surface diffusion. Finally, the growth conditions responsible for the formation of $\delta$- and $\epsilon$-NbN, have been verified using RBS and depth dependent XPS measurements.
Superconducting single photon detectors were fabricated and characterised for $4$ and $22$nm thick optimised NbN films. The electro-optical characterisation of the devices revealed an extension of the photon induced resistive region of $270\pm10$nm for the thin SSPD and $30\pm8$nm for the thick detector.
By varying the normally incident photon number and measuring the detected photon events, the single photon character of both devices was proven.
The maximum detection efficiency was determined to be $0.004\pm0.001\%$ for the thick detector and $21\pm2\%$ for the thin detector, following excitation with $950$nm photons.
The origin of the negligibly low dark count rate for the thicker detector, operated at $4.3\pm0.1$K, was attributed to hotspot quenching with the $22$nm thick film acting as a heat sink.
The evanescent coupling of such SSPDs with waveguides is likely to significantly improve these performance metrics, providing strong potential for the development of integrated single photon quantum optics using semiconductors.

We gratefully acknowledge D. Sahin, A. Fiore (TU Eindhoven) and K. Berggren, F. Najafi (MIT) and R. Hadfield (Heritot-Watt) for useful discussions, EAG Labs for film characterisation via RBS / XPS and the BMBF via QuaHL-Rep Project number 01BQ1036. 

\bibliography{Papers}

\end{document}